\documentclass[fleqn,usenatbib,useAMS]{mnras}
\usepackage{graphicx}   
\usepackage{amsmath}    
\usepackage{amssymb}    
\usepackage{multicol}        
\usepackage{bm}     
\usepackage{pdflscape}  
\usepackage{longtable}

\usepackage{threeparttablex}

\usepackage[T1]{fontenc}
\usepackage{ae,aecompl}

\title[Statistic of Width of Supra-Arcade Downflows]{Statistical Investigation of the Widths of Supra-arcade Downflows Observed During a Solar Flare}

\author[Tan et al.]{
Guangyu Tan$^{1,3}$,
Yijun Hou$^{2,3,4,5}$\thanks{Corresponding author, \href{mailto:yijunhou@nao.cas.cn}{yijunhou@nao.cas.cn}},
and Hui Tian$^{1,2,4}$
\\
$^{1}$School of Earth and Space Sciences, Peking University, Beijing 100871, China
\\
$^{2}$National Astronomical Observatories, Chinese Academy of Sciences, Beijing, 100101, China
\\
$^{3}$Yunnan Key Laboratory of the Solar physics and Space Science, Kunming, 650216, China
\\
$^{4}$Key Laboratory of Solar Activity and Space Weather, National Space Science Center, Chinese Academy of Sciences, Beijing 100190, China
\\
$^{5}$School of Astronomy and Space Science, University of Chinese Academy of Sciences, Beijing 100049, China
}
\pubyear{2023}
\begin{document}
\label{firstpage}
\pagerange{\pageref{firstpage}--\pageref{lastpage}}
\maketitle

\begin{abstract}
Supra-arcade downflows (SADs) are dark voids descending towards the post-reconnection flare loops and exhibit obvious variation in properties like width. However, due to the lack of further statistical studies, the mechanism behind such variations hitherto remains elusive. Here we statistically investigated widths of 81 SADs observed in one flare by the {\it Solar Dynamics Observatory (SDO)}. For each of SADs, six moments were selected with equal time intervals to measure their widths at different stages of their evolution. It is found that most SADs show a roughly monotonous width decrease during their descents, while some SADs with small initial widths can have complex evolutions. 3D reconstruction results based on {\it SDO} and {\it Solar Terrestrial Relations Observatory Ahead (STEREO-A)} images and thermal properties analysis reveal that differences in magnetic and plasma environments may result in that SADs in the north are overall wider than those in the south. Additionally, correlation analysis between the width and other parameters of SADs was further conducted and revealed that: (1) SADs with different initial widths show no significant differences in their temperature and density evolution characteristics; (2) SADs with small initial widths usually appear in lower heights, where more frequent collisions between SADs could lead to their intermittent acceleration, width increment, and curved trajectories. These results indicate that SADs with different initial widths are produced the same way while different environments (magnetic field or plasma) could affect their subsequent width evolutions.
\end{abstract}




\section{Introduction} \label{sec:intro}
Solar flares are one of the most violent events in the solar atmosphere, which is caused by magnetic reconnection. Supra-arcade fans are diffuse regions above flare arcades with $\sim$10 MK plasma \citep{1999ApJ...519L..93M,2000SoPh..195..381M,2014ApJ...786...95H} and are often observed at extreme ultraviolet (EUV) and soft X-ray (SXR) wavelengths \citep{2018ApJ...866...29F}. Dark tadpole-like voids are usually observed to descend through these fans and called supra-arcade downflows (SADs) \citep{1999ApJ...519L..93M}. They were first observed by the Yohkoh Soft X-ray Telescope \citep[SXT;][]{2000SoPh..195..381M,2007A&A...475..333K}, and always found in the early decay phases of long-term eruptive flares that produce coronal mass ejections (CMEs) \citep{1999ApJ...519L..93M, 2002ApJ...579..874S, 2003SoPh..217..247I, 2010ApJ...722..329S}. However, \citet{2004ApJ...605L..77A} and \citet{2007A&A...475..333K} found that SADs can also happen in the impulsive phase of solar flares.

SADs have a typical lifetime of a few to ten minutes, and their velocities are about 20--500 km s$^{-1}$ \citep{2011ApJ...730...98S,2011A&A...527L...5M,2022MNRAS.tmp.2292T,2022ApJ...933...15X}. SADs are filled with low-density plasma \citep{2003SoPh..217..247I,2012ApJ...747L..40S}, and most SADs are cooler than the surrounding fan plasma according to the results of differential emission measure (DEM) analysis \citep{2014ApJ...786...95H,2017ApJ...836...55R,2020ApJ...898...88X}. Recently, \citet{2022FrASS...9.2607Z} found that radio observations could be helpful to detect SADs through modeling, which will provide a new method to calculate the temperature and density of SADs.

SADs are also closely associated with other phenomena or physical processes. For example, recently, \citet{2021Innov...200083S} found obvious soft X-ray emission enhancement where SADs collide with the flare loop, indicating heating of local plasma to 10-20 MK. Similar phenomenon was also reported by \citet{2022ApJ...941..158A}, who proposed that SADs could cause quasi-periodic pulsation observed during flare gradual phase and play an important role in transporting flaring energy to the low atmosphere. Furthermore, \citet{2019PhRvL.123c5102S} found that SADs can cause oscillations in plasma upflows through vortex shedding, which is the first time this mechanism has been observed on the Sun. In addition, SADs can result in the temperature increase of regions with SADs passby, which could be the adiabatic compression in front of SADs \citep{2017ApJ...836...55R,2020ApJ...898...88X,2021ApJ...915..124L,2022MNRAS.tmp.2292T} and viscous heating \citep{2017ApJ...836...55R}. Three-dimensional (3D) magnetohydrodynamic (MHD) simulation also suggested that adiabatic compression is important for plasma heating in large scale current sheet regions in the eruption\citep{2019ApJ...887..103R}.

Although SADs play an important role in the energy transport in the solar atmosphere, their formation mechanism is still in dispute. In order to explore the formation process of SADs, there are many observational \citep{1999ApJ...519L..93M, 2009ApJ...697.1569M, 2011ApJ...742...92W, 2014ApJ...786...95H, 2017A&A...606A..84C} and simulation \citep{2013ApJ...775L..14C,2014ApJ...796L..29G, 2014ApJ...796...27I,2015ApJ...807....6C, 2016ApJ...832...74Z} studies. At first, \citet{1999ApJ...519L..93M} proposed that SADs are the cross-sections of evacuated flux tubes. And with high-resolution data from the Solar Dynamics Observatory \citep[SDO;][]{2012SoPh..275....3P}/Atmospheric Imaging Assembly \citep[AIA;][]{2012SoPh..275...17L}, a further explanation was proposed by \citet{2012ApJ...747L..40S} that SADs are the wakes behind the shrinking flare loops, which is consistent with the results based on three-dimensional bursty reconnection model \citep{2006ApJ...642.1177L,2011ApJ...730...90G}. The SAD model developed by \citet{2011A&A...527L...5M} reported that SADs are a consequence of the shocks happened in intermittent, bursty magnetic reconnection. \citet{2013ApJ...775L..14C} pointed out the wakes would be filled too quickly by the surrounding denser and hot plasma, and proposed that SADs are low-density outflow jets produced by reconnection penetrating into the denser flare arcades but not in the supra-arcade fans. On the other hand, some models have demonstrated that instability also contributes to the formation of SADs. Rayleigh-Taylor instability (RTI) between the reconnection outflow and the denser plasma sheet can produce the SADs \citep{2014ApJ...796L..29G,2014ApJ...796...27I}. Recently, \citet{2022NatAs.tmp...29S} proposed that RTI and Richtmyer-Meshkov instability (RMI) happening in the turbulent interface region between post-reconnection flare arcades and plasma sheet can also produce SADs.

Despite many observational interpretations and simulated models proposed in the past, nature of some observational features of SADs still can not be successfully explained to date. And the variation of SAD width is among the most elusive one. Developing an automatically tracking algorithm, \citet{2022ApJ...933...15X} performed a statistical study of SADs detected from multiple solar flares and found that the width of SADs is about 1.2--20 arcsec and has log-normal distribution characteristic showing good consistency with the result reported by \citet{2011ApJ...735L...6M}. SADs are usually observed when the supra-arcade fans is face-on, i.e., the line of sight (LOS) is perpendicular to the arcades' axis, and high temporal resolution EUV observations reveal that the appearance of SADs would change with evolution of the fan, e.g., from wide to thin and from short to long \citep{2010ApJ...722..329S,2011ApJ...742...92W}. Although some papers have investigated width of SADs, due to the lack of further statistical study of width variation of SADs (especially observed in one single flare), several open questions still remain: What are the distribution and evolution characteristics of widths of SADs during one flare? What are the key factors determining the width of these SADs as well as their evolution characteristics? What are the differences of other parameters of these SADs with different widths? Do the SADs with different widths have the same formation mechanism?

Insights into these questions are necessary to understand the width variation and formation mechanisms of SADs and call for a comprehensive statistical investigation on the width of SADs observed during one flare. In this paper, we employed the SAD sample constructed by \citet{2022MNRAS.tmp.2292T} based on observations from the SDO/AIA, and statistically investigated the distribution and evolution characteristics of widths of SADs appearing in one flare. In addition, correlation analysis between the width and other physical parameters of SADs is also conducted. The remainder of this paper is organized as follows. Section \ref{sec:Obs} presents the observations and data analysis method used in this work. Section \ref{sec:Res} shows results of SAD width distributions and correlation analysis with other parameters. In Section \ref{sec:Con}, we briefly summarize our findings.

\section{Observations and data analysis} \label{sec:Obs}
\subsection{Observations} \label{sec:Obss}
The SAD sample analyzed here was constructed in our recent work \citep{2022MNRAS.tmp.2292T} and based on the SDO/AIA observations. The AIA telescopes cover a full-disk field of view (FOV) up to $\sim$ 1.30 $R_{\odot}$, thus capable of observing flares near the solar limb \citep{2012SoPh..275...17L}. The 6 EUV passbands (94, 131, 171, 193, 211, and 335 \AA) on AIA are sensitive to corona temperature (from $10^{5.5}$ to $10^{7.5}$ K) and have a high temporal cadence of 12 s and angular resolution of  $\sim$0.6\arcsec. The AIA 131 \AA\ passband is mainly contributed by Fe {\sc xxi} and the plasma temperature is about 10 MK in which the SADs always are observed. We used the {\tt aia\_prep} routine which is available in the Solar Software \citep[SSW;][]{1998SoPh..182..497F} package to rotate and align AIA images.

To reconstruct 3D structure of flare loops below the SADs, we also employ observations made by the Extreme Ultraviolet Imager \citep[EUVI;][]{2004SPIE.5171..111W} telescope on board the Solar Terrestrial Relations Observatory Ahead (STEREO-A). The EUVI instrument is part of the Sun Earth Connection Coronal and Heliospheric Investigation instrument \citep[SECCHI;][]{2008SSRv..136...67H}, and can observe the low corona like 195 \AA\ passband to 1.7 $R_{\odot}$ with spatial resolution of $\sim$1.6\arcsec. The EUVI data were processed by using the {\tt secchi-prep} routine available in SSW.

\subsection{Identification of SADs} \label{subsec:identification}
\begin{figure*}
\centering
\includegraphics [width=0.75\textwidth]{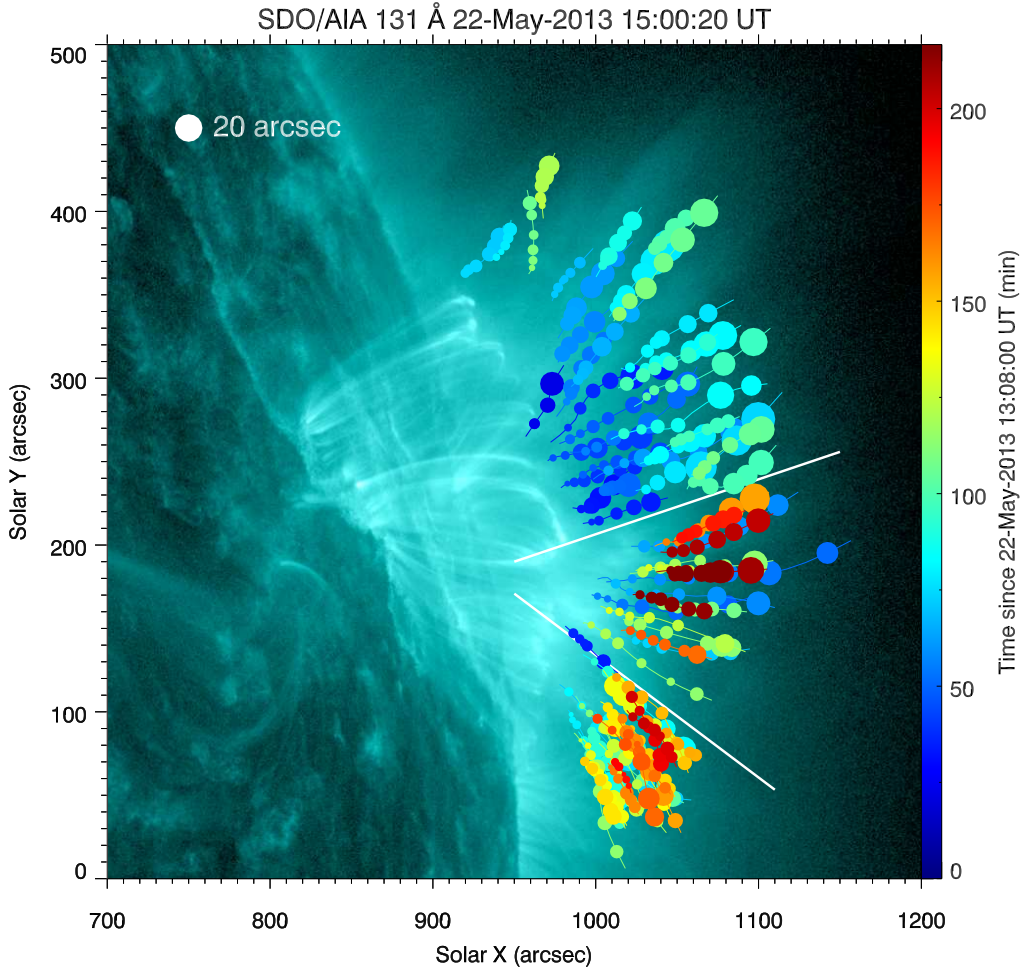}
\caption{81 SADs identified in the hot flare region. The trajectories of SADs are approximated by curves with different colors corresponding to their appearance times. Six circles with different radii in each curve represent the evolutive width of one SAD during its descent. Two white lines divide the whole area into north region, middle region, and south region, where appearance times of SADs exhibit different features.}
\label{fig:fig1}
\end{figure*}

As introduced in \citet{2022MNRAS.tmp.2292T}, 81 SADs were identified according to the frame-by-frame check of AIA 131 \AA\ images. More details can be found in that paper. In Figure \ref{fig:fig1}, we plot trajectories of all the 81 SADs on one 131 {\AA} image with different colors according to their appearance times.

\subsection{Width Measurements} \label{subsec:width}
\begin{figure*}
\centering
\includegraphics [width=0.7\textwidth]{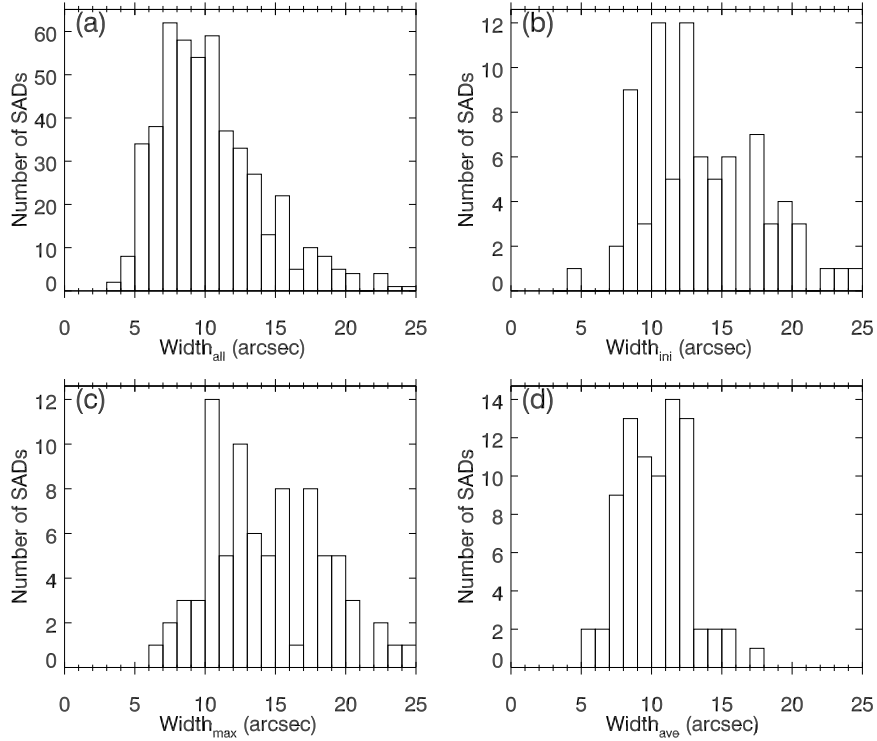}
\caption{Width distributions of 81 SADs. \textbf{a}: Distribution of all widths at the six moments of each SAD during its descent. \textbf{b}: Distribution of the initial width of each SAD during its descent. \textbf{c}: Distribution of the max width of each SAD during its descent. \textbf{d}: Distribution of the average width of each SAD during its descent.}
\label{fig:fig2}
\end{figure*}

Because the morphological outline of SADs is not obvious when they first appeared and finally disappeared, parameter measurement at the beginning and end times will inevitably cause large errors. To avoid this kind of measurement errors, here we define the start and end time of SADs as one minute after their appearance and before their disappearance, respectively. Then, period between these two time points is divided into 5 equal parts, and we obtain six moments: $\text{t}_{0}$, $\text{t}_{1}$, $\text{t}_{2}$, $\text{t}_{3}$, $\text{t}_{4}$, $\text{t}_{5}$ (thus six sites), to measure various physical parameters of SADs, including width, temperature, density, velocity, and trajectory curvature. At each moment, SAD width is uniformly measured at the site 8 arcsecond behind the front of SAD and defined as the spatial scale of SAD along the direction perpendicular to its trajectory. In Figure \ref{fig:fig1}, each SAD has six different widths during its descent, which are marked by circles with different radii. As a result, the initial ($w_{ini}$), maximum ($w_{max}$), and average widths ($w_{ave}$) of each SAD can be calculated, and their distributions of 81 SADs are shown in Figure \ref{fig:fig2}.

\subsection{DEM Analysis} \label{subsec:dem}
DEM analysis is also performed to study the temperature and density characteristics of SADs. The codes we used to perform the DEM analysis were written by \citet{2015ApJ...807..143C} and improved by \citet{2018ApJ...856L..17S}. The specific calculation method is described in our previous article \citep{2022MNRAS.tmp.2292T}. In order to compare the thermal parameters more accurately, after determining the position of the front of SADs, we uniformly defined the position 8 arcsecond behind the front as the site of body for temperature and density measurements.

\subsection{Velocity and Trajectory Curvature Calculations} \label{subsec:vel_cur}
Velocities and trajectory curvatures at the 6 moments are also calculated for each SAD. For each moment, we calculated an average velocity within one minute before and after this moment. To represent curvature of the whole trajectory, we define a variate $Cur$ as the sum of the absolute value of the curvature values of the six points and the maximum curvature value of the trajectory:

\begin{equation}
\centering
Cur=\sum_{i=0}^{5} \left | cur_{i} \right |  +{\rm max}(\left | cur \right |  )
\label{eq:cur}
\end{equation}
where $cur_{i}=\frac{\bigtriangleup \alpha_{i} }{\bigtriangleup s_{i}}$ is the curvature of the i-th point. $\bigtriangleup \alpha_{i}$ represents the change value of tangent angle of the trajectory within 12 s before and after the i-th point, and $\bigtriangleup s_{i}$ represents length of the trajectory within 12 s before and after the i-th point.

\section{Result and Discussion} \label{sec:Res}
In this section, we show the results of SAD width distributions and correlation analysis between width and other parameters of SADs. For the correlation analysis, SADs are respectively classified into two different groups from the following perspectives: (1) According to the initial width ($w_{ini}$), 15 SADs with the largest $w_{ini}$ and 15 SADs with the smallest $w_{ini}$ in the north region are selected; (2) According to the evolution characteristic of width, SADs are divided into 2 groups, one of which perform roughly monotonous decrease of width during the descent (SAD$_{Wdec}$) while the other one perform complicated evolution of width (SAD$_{Wcom}$); (3) According to the evolution characteristic of body and front temperature, SADs are divided into 2 groups, one of which increase gradually ((T$_4$+T$_5$)/2 \textless\ 2T$_0$, SAD$_{Tgra}$) while the other one increase impulsively ((T$_4$+T$_5$)/2 $\geq$ 2T$_0$, SAD$_{Timp}$); (4) According to the evolution characteristic of body and front density, SADs are divided into 2 groups, one of which increase gradually ((N$_4$+N$_5$)/2 \textless\ 2N$_0$, SAD$_{Ngra}$) while the other one increase impulsively ((N$_4$+N$_5$)/2 $\geq$ 2N$_0$, SAD$_{Nimp}$); (5) According to the evolution characteristic of velocity, SADs are divided into 2 groups, one of which perform great decrease of velocity at middle time ((V$_2$+V$_3$)/2 \textless\ 0.5V$_{0}$, SAD$_{Vdec}$) while the other one have a significant acceleration ((V$_2$+V$_3$)/2 \textgreater\ V$_{0}$, SAD$_{Vacc}$); (6) According to the trajectory curvature, SADs are divided into 2 groups, one of which is curved ($Cur$ $\geq 0.4$, SAD$_{Cur}$) while the other one is nearly straight ($Cur$ \textless 0.4, SAD$_{Str}$). In Table \ref{tab:SADs}, we show the initial and maximum width of the 81 SADs and their various classifications mentioned above.

\subsection{Overall Distribution} \label{subsec:ove}
Figure \ref{fig:fig1} shows the spatial distribution of 81 SADs in the flare region. It is revealed that as the flare evolved, the region where SADs mainly appear moved from the north part to the south part \citep{2022MNRAS.tmp.2292T}. According to the appearance time, we divided the whole region into three parts: the north region where SADs are concentrated in early stage; the south region where SADs are concentrated in late stage; the middle region where the appearance times of SADs are relatively scattered. Moreover, one can see that the SADs in the north region have larger widths than those in the south region overall. We suggest that such difference could be caused by different magnetic environments.

\begin{figure*}
\centering
\includegraphics [width=0.99\textwidth]{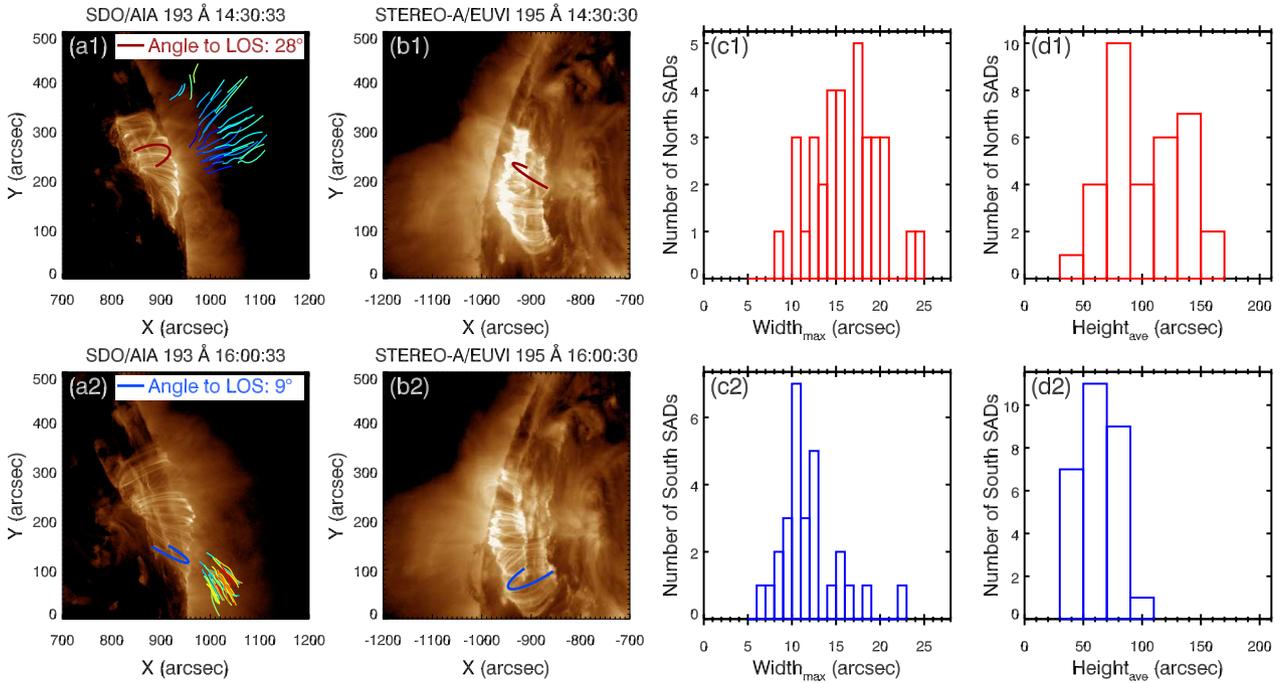}
\caption{Differences in magnetic environments, width distributions, and average height distributions of SADs in the north region and south region. \textbf{a1-a2 and b1-b2}: SDO/AIA 193 and STEREO-A/EUVI 195 \AA\ images showing flare loops below the SADs. Two loops selected in the north and south regions are marked by red and blue curves, respectively. Their angles to LOS are calculated through the 3D ellipse loop fitting. \textbf{c1-c2 and d1-d2}: Statistical distributions of the width and average height of SADs observed in the north and south regions, respectively.}
\label{fig:fig3}
\end{figure*}

In Figure \ref{fig:fig2}, we show statistical distribution of widths for 81 SADs, including all the calculated widths ($w_{all}$), initial width ($w_{ini}$), maximum width ($w_{max}$), and average width ($w_{ave}$) during the descent of SADs. As shown in Figure \ref{fig:fig2}(a), $w_{all}$ present an approximate log-normal distribution, which is consistent with the results of \citet{2022ApJ...933...15X}. Similar log-normal distribution was also found in width of plasmoids during a CME or late time dissipation rate during RMI in fluid mechanics\citep{2012PhFl...24g4105W,2013ApJ...771L..14G,2020A&A...644A.158P}. About the nature of the log-normal distribution of SAD width, \citet{2011ApJ...735L...6M} provided a possible scenario with minimal assumptions, in which SADs forming through patchy reconnection grow at a rate proportional to the size of the patch. However, in Figures \ref{fig:fig2}(b)-(d), distributions of $w_{ini}$, $w_{max}$ and $w_{ave}$ show a double-peak structure. We suggest it could be caused by the width difference of SADs in the north and south regions (as shown in Figure \ref{fig:fig1}).

\subsection{Relationship between maximum width and magnetic field configuration} \label{subsec:maxwidit}
To check if the double-peak structure shown in Figures \ref{fig:fig2}(b)-(d) are caused by the width difference of SADs in the north and south regions, we further analyzed respective width distributions of SADs in the two regions. The distribution of $w_{max}$ of SADs in the north region is shown in Figure \ref{fig:fig3}(c1) and exhibits an average value of $\sim$ 16.2 arcsecond. Figure \ref{fig:fig3}(c2) shows the distribution of $w_{max}$ of SADs in the south with an average value of $\sim$ 11.9 arcsecond. Obviously, the northern SADs have larger widths than those in the south region. We speculate that it is the different magnetic environments that result in the difference of observed widths of SADs in these two regions through projection effect. To verify this point, we combined observations from two perspectives of SDO and STEREO-A to reconstruct 3D structures of flare loops below SADs in the north and south regions. Detailed description about the reconstruction method can be found in \citet{2021ApJ...911....4L}. 3D structures of two loops in the north and south regions are reconstructed for the two moments when SADs are concentrated in the north and south regions, respectively. It is revealed that the angle of the flare loop in the north region to the LOS is 28$^{\circ}$ around 14:30 UT (Figures \ref{fig:fig3}(a1) and (b1)), and the angle of the flare loop in the south region to the LOS is 9$^{\circ}$ around 16:00 UT (Figures \ref{fig:fig3}(a2) and (b2), showing remarkable difference. As a result, it is reasonable to say that the magnetic environments in the north and south regions are different, which then affects the observed widths of SADs there.

\begin{figure*}
\centering
\includegraphics [width=0.85\textwidth]{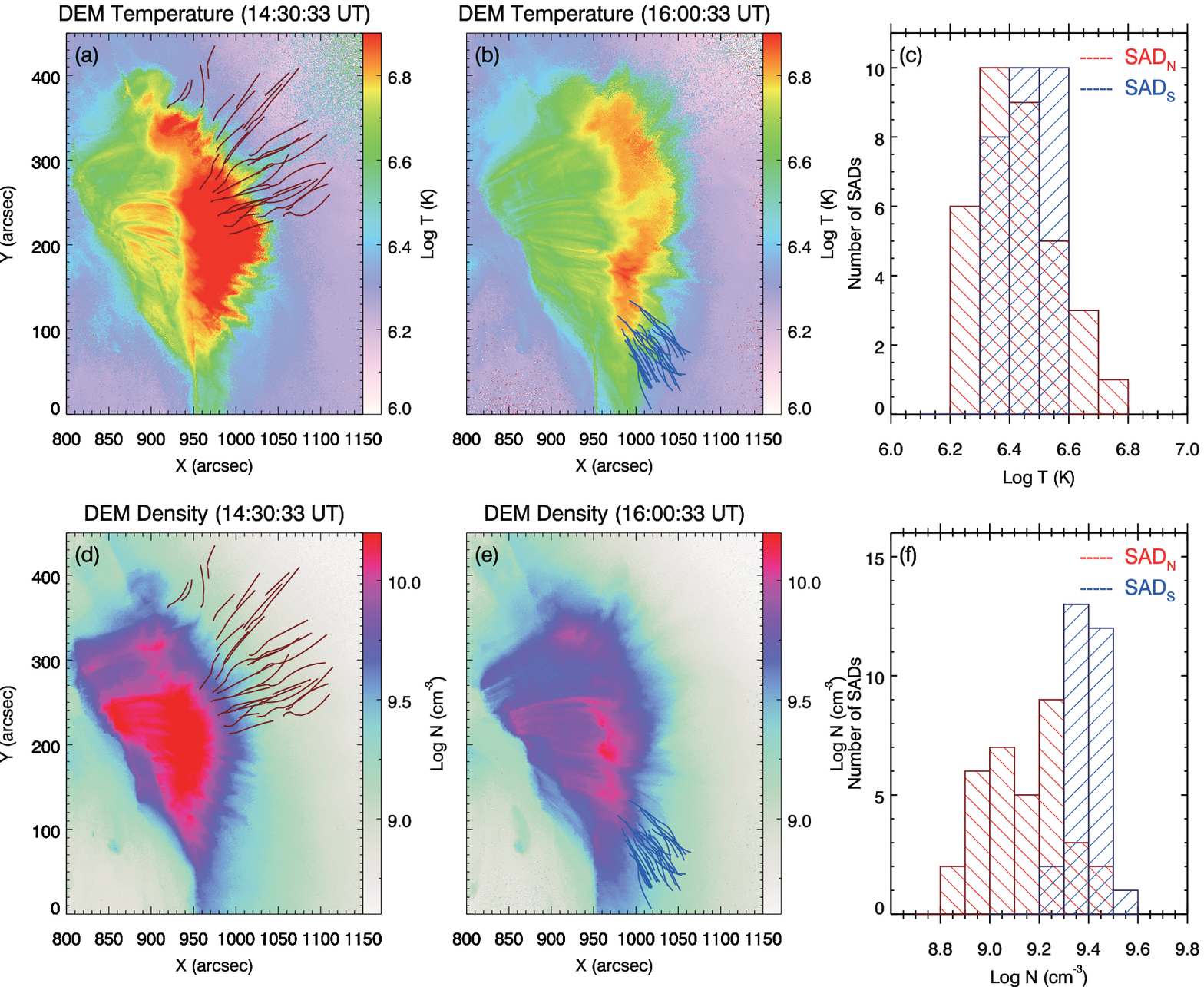}
\caption{Properties of plasma surrounding SADs in the north and south regions. \textbf{a,b}: Temperature maps at two moments when SADs are concentrated in the north and south regions, respectively. \textbf{c}: Statistical distribution of temperature of SADs' ambient plasma in the north and south regions. \textbf{d,e}: Similar to \textbf{a,b}, but for density maps. \textbf{f}: Similar to \textbf{c}, but for statistical distribution of density.}
\label{fig:figadd}
\end{figure*}

Figures \ref{fig:fig3}(d1-d2) show the average height distributions of the SADs in the north and south regions. In our study, the average height of SADs is defined as the average of the heights at six moments during the descent. And the heights we measured all refer to the radial distances between the SAD and the solar limb. The result reveals that the average height of SADs in the north region is relatively higher, hinting that plasma environments (e.g., temperature and density) of SADs in the north and south regions could be different.

\subsection{Relationship between maximum width and thermal properties} \label{subsec:demnt}
In order to find out whether the properties of plasma surrounding SADs in the north and south regions is related to the significant difference of their $w_{max}$, we have investigated the temperature and density of fan plasma surrounding SADs. In Figures \ref{fig:figadd}(a,b) and (d,e), we show temperature maps and density maps at two moments when SADs are concentrated in the north and south regions, respectively. For each SAD, we measure the temperature and density of six SAD sites at the time 3 minutes before its arriving moment and then obtain an average temperature and density value. Figure \ref{fig:figadd}(c) reveals that there is no obvious difference of ambient plasma temperature between the SADs with different widths in the north and south regions. But the plasma surrounding SADs with smaller widths in the south region have larger density than that of SADs with larger widths in the north region (Figure \ref{fig:figadd}(f)). Combining the fact shown in Figures \ref{fig:fig3}(d1-d2) that the average height of SADs in the north region is relatively higher than that in the south region, we propose that due to lower occurrence site of magnetic reconnection, the SADs in the south region appear in a relatively lower altitude thus have a plasma environment with higher density, which tends to breed SADs with small widths.

\begin{figure*}
\centering
\includegraphics [width=0.78\textwidth]{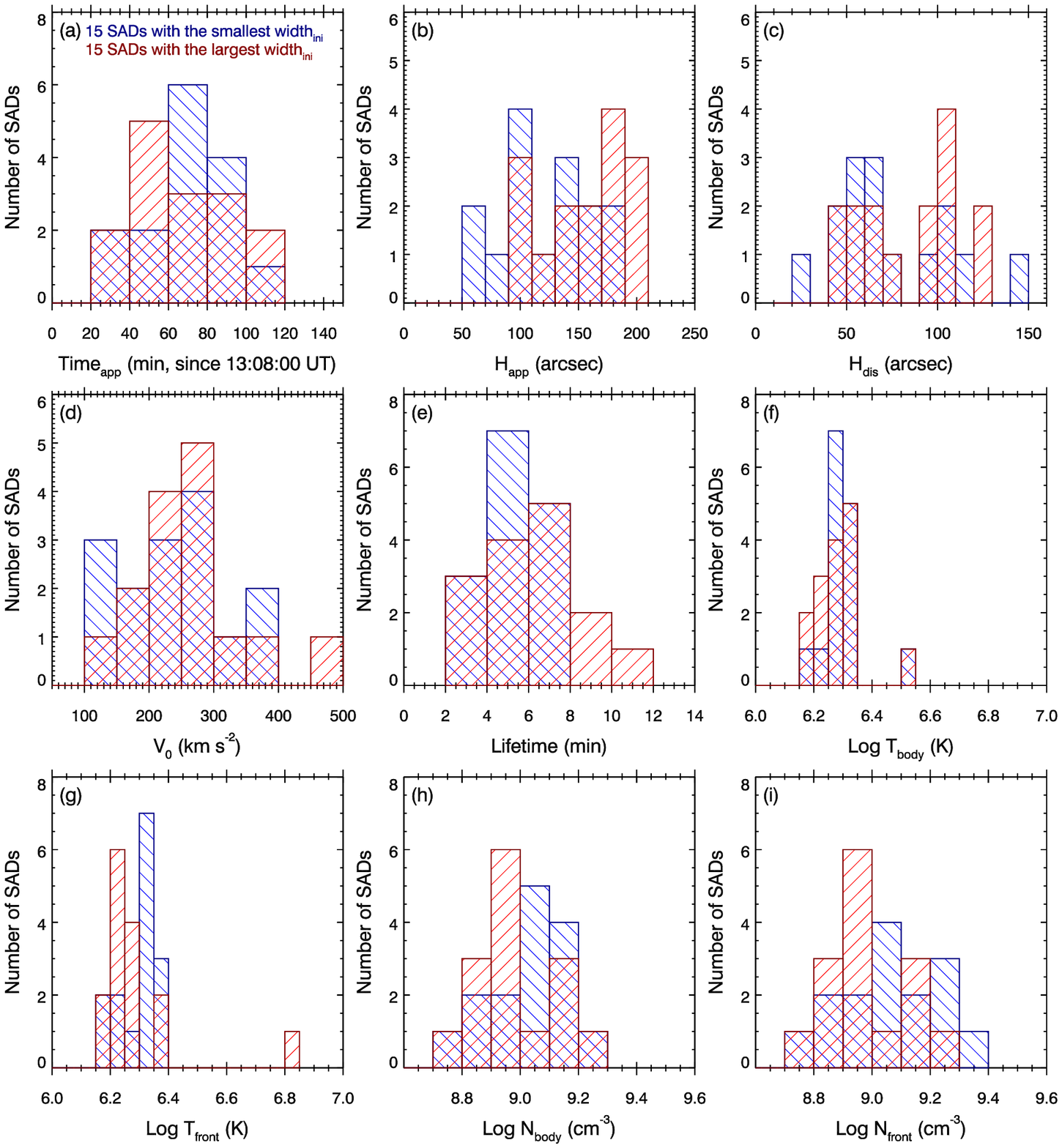}
\caption{Other parameter distributions of two groups of SADs with the largest and smallest initial width in the north region. For the two groups of SADs, distributions of their appearance times ($\text{Time}_{app}$ at $\text{t}_{0}$), appearance heights ($\text{H}_{app}$ at $\text{t}_{0}$), disappearance heights ($\text{H}_{dis}$ at $\text{t}_{5}$), velocities($\text{V}_{0}$ at $\text{t}_{0}$), lifetimes, temperatures, and densities are represented by red and blue histograms, respectively.}
\label{fig:fig4}
\end{figure*}

It is worth noting that the comparative analysis of plasma properties is conducted here to the two groups of SADs with different observed widths located in two different regions, where the magnetic environments and heights of the SADs have been proven to be different. As a result, the different widths of SADs in these two regions could result from a combination of magnetic and plasma factors, among which it is difficult to determine the dominant one. For an unbiased evaluation of influence of plasma environment on the evolution of SADs' width, in the future, we need to select SADs with similar appearance width, height, velocity, and magnetic environment. Then we can investigate how the width of these SADs would evolve when they enter into lower atmospheres with different plasma environments during their descents.

\subsection{Other parameter characteristics of two groups of SADs with the largest and smallest initial widths} \label{subsec:widittwo}
Besides external factors like the magnetic environment and plasma environment discussed above, variation of SAD width could also be essentially caused by their different formation mechanisms. If the SADs with different widths are produced by different mechanisms, their initial widths, other physical parameters, and their subsequent evolutions would naturally show obvious differences. As a result, in the following subsections, correlation analysis between the initial width, other parameters, and their evolutions was further conducted to explore other potential factors determining the width variation.

\begin{figure*}
\centering
\includegraphics [width=0.85\textwidth]{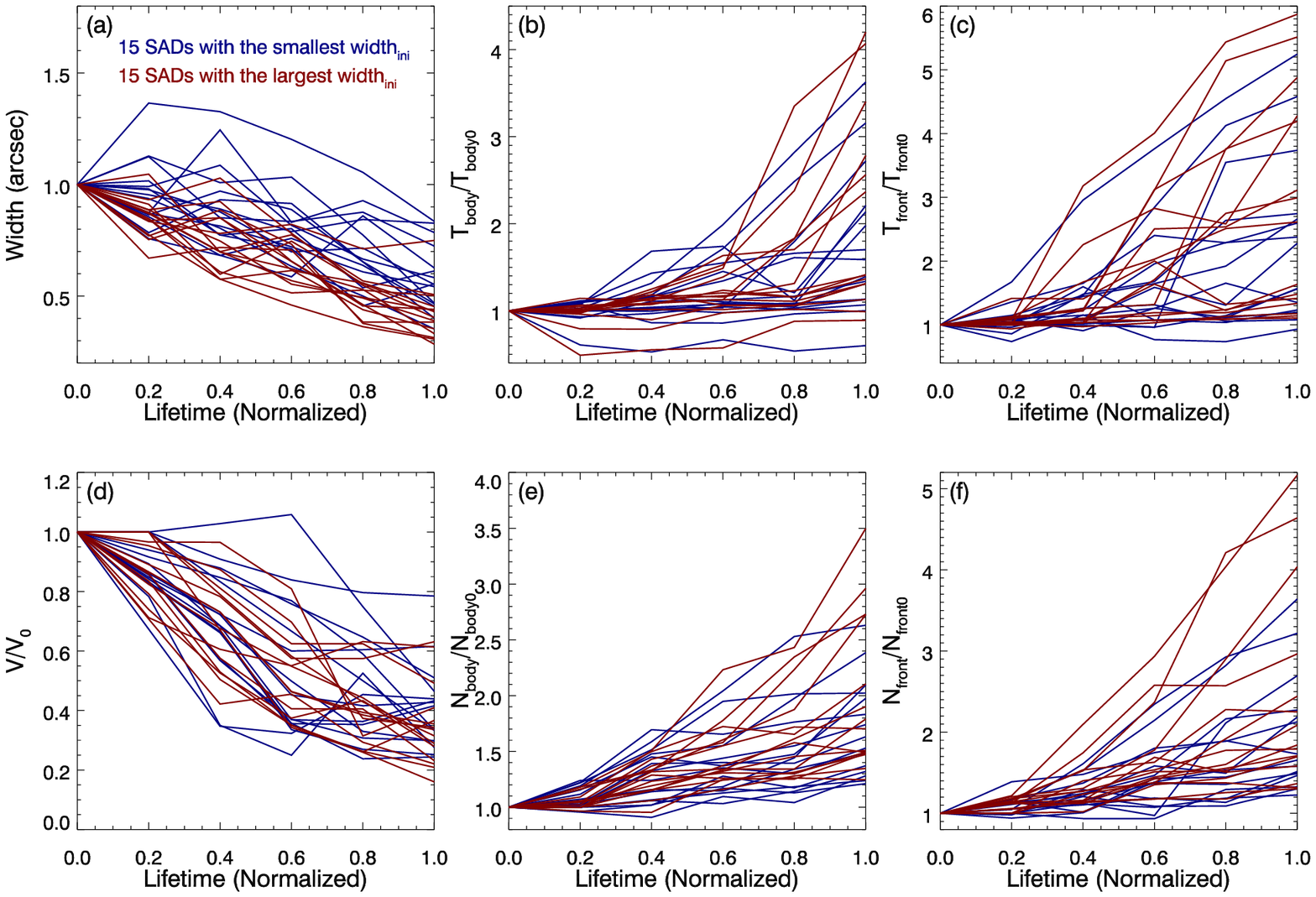}
\caption{Other parameter evolutions of two groups of SADs with the largest and smallest initial width in the north region. For the two groups of SADs, evolutions of their widths, temperatures, velocities, and densities are represented by red and blue broken line, respectively. }
\label{fig:fig5}
\end{figure*}

Firstly, we selected two groups of SADs with significantly different initial widths to check if their other parameters, like appearance time, appearance height, disappearance height, velocity at $\text{t}_{0}$ moment (V$_{0}$), lifetime, temperature and density at $\text{t}_{0}$ moment, show obvious differences. In order to eliminate the interference caused by different magnetic field configurations in the north and south regions, we selected 15 SADs with the largest initial widths and 15 ones with the smallest initial widths in the north region for parameter comparison. As shown in Figures \ref{fig:fig4}(a), (c), and (d), the two groups of SADs show no significant differences in their appearance time, disappearance heights, and initial projective velocities. From panels (b) and (e), one can see that SADs with larger initial widths naturally have larger appearance heights and lifetimes than those with smaller initial widths. Panels (f)-(i) show that the SADs with larger widths tend to have lower temperatures and densities both in body and front, which could result from their higher appearance heights.

\begin{figure*}
\centering
\includegraphics [width=0.99\textwidth]{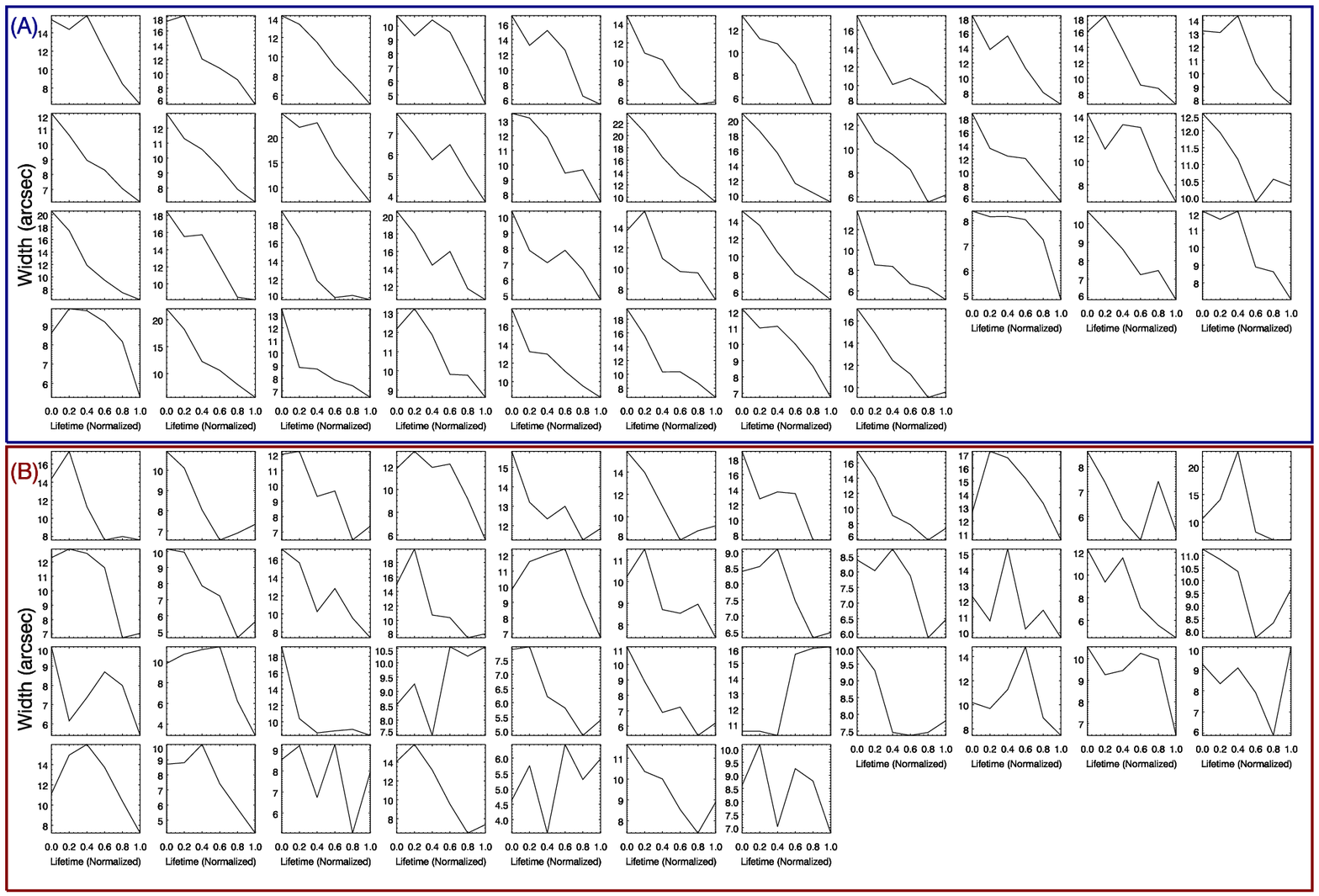}
\caption{Different width evolution properties of the 81 SADs. These SADs are classified into two types: SADs with roughly monotonous decreasing widths during the descent (A, SAD$_{Wdec}$); SADs with widths evolving complicatedly during the descent (B, SAD$_{Wcom}$).}
\label{fig:fig6}
\end{figure*}

After comparing the distributions of different parameters of the two groups of SADs, we further check their evolution characteristics in Figure \ref{fig:fig5}. It is revealed that for most SADs with large initial widths, their widths decrease monotonously during the descent while widths of some SADs with small initial widths show complicated evolution (panel (a)). Panels (b,c) and (e,f) show that the two groups of SADs have similar temperature and density evolution characteristics, both in their front and body. And panel (d) reveals that some SADs with small initial widths could show complicated descent with obvious accelerations while most SADs with large initial widths show roughly monotonous decelerations.

After comparing parameter characteristics of the two groups of SADs with distinctly different initial widths, according to the classifications based on the evolution features of different parameters mentioned at the begin of this section, we further analyzed width distributions of the different types of SADs.

\begin{figure*}
\centering
\includegraphics [width=0.9\textwidth]{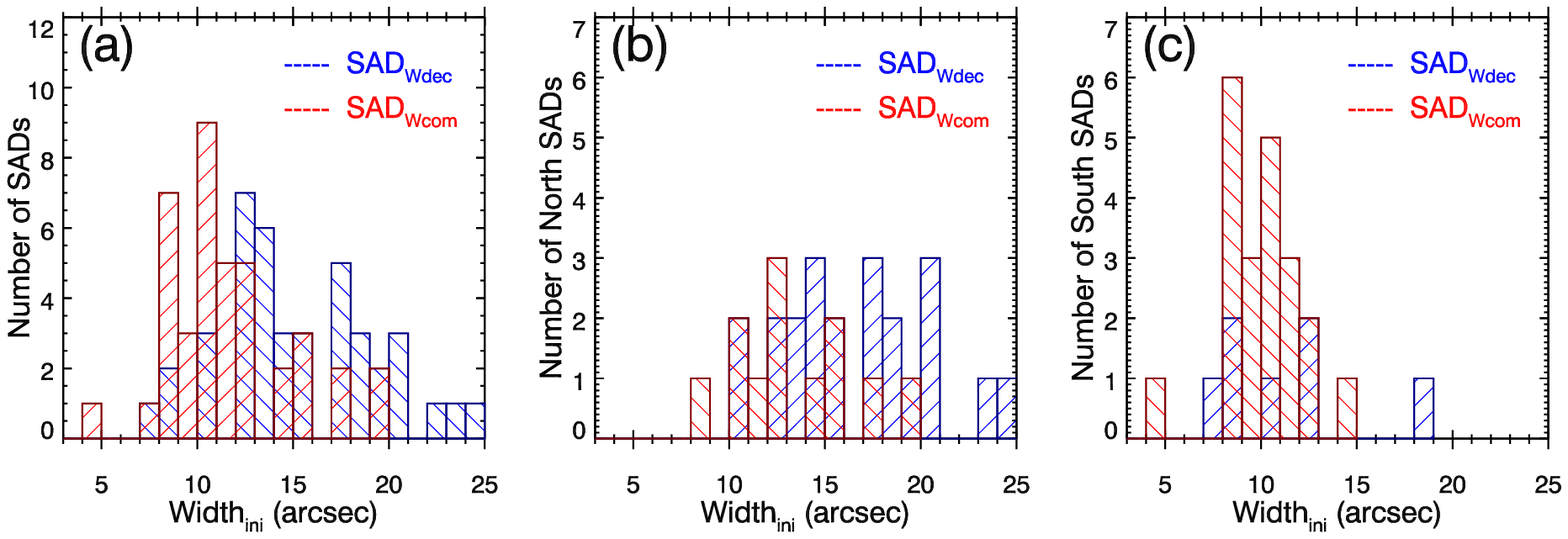}
\caption{Width distributions of SADs with different width evolution properties. According to the classification defined in Figure \ref{fig:fig6}, width distributions of two types of SADs (SAD$_{Wdec}$, SAD$_{Wcom}$) in the whole region, in the north region, and in the south region are shown in \textbf{a}, \textbf{b}, and \textbf{c}, respectively.}
\label{fig:fig7}
\end{figure*}

\subsection{Width distributions of SADs with different width evolution properties} \label{subsec:widit}
Figure \ref{fig:fig6} exhibits the evolution of widths of 81 SADs, and they are divided into 2 groups: SAD$_{Wdec}$ and SAD$_{Wcom}$ as mentioned at the beginning of this Section. To be specific, the SADs with $\geq$2 moments showing width increment or 1 moment with a $\geq$20\% width increase compared to its previous moment are classified as SAD$_{Wcom}$. And the other SADs are identified as SAD$_{Wdec}$. According to this classification, initial width distributions of the two types of SADs are shown in Figure \ref{fig:fig7}(a). To avoid effect of the magnetic environment factor, SADs in the north and south regions are further shown in panels (b) and (c), respectively. SAD$_{Wdec}$ tend to have larger initial widths than SAD$_{Wcom}$, which is consistent with Figure \ref{fig:fig5}(a). We suggest that the SADs with large initial widths tend to appear at a higher height, where the magnetic pressure is relatively small. And it is nature that these SADs would show obvious decrease in their spatial scale during the descent to lower atmosphere with stronger magnetic fields. As for the SAD$_{Wcom}$, they are found to have smaller initial widths and appear at a lower height (Figure \ref{fig:fig10_check}(a)), where collision or mergence would occur more frequently between SADs and lead to the width increment. Similar processes have also been reported in \citet{2012ApJ...747L..40S} that two SADs can merge and become a larger descending void. In addition, it should be noted that some SADs will split during the descend\citep{2014ApJ...796...27I}. During our identification of SADs in this work, when one SAD splits into two parts, we would continue to track the relatively larger part and measure its width as the SAD's width in the following time. As a result, the splitting of SADs will naturally reduce their measured widths, which could result in the rapid decrease of widths during the evolution of some SADs.

\begin{figure*}
\centering
\includegraphics [width=0.98\textwidth]{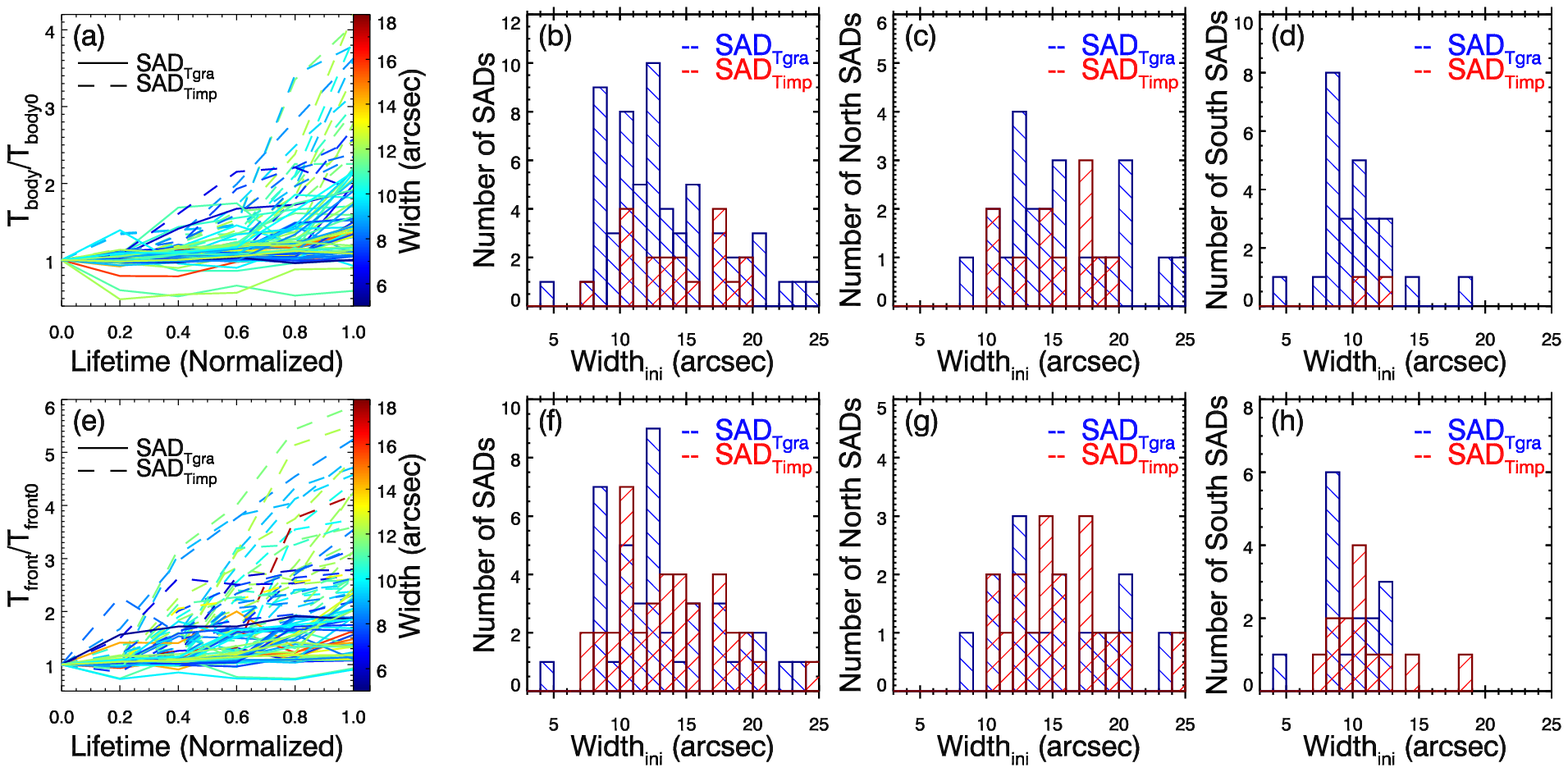}
\caption{Width distributions of SADs with different temperature evolution properties. \textbf{a,e}: Evolutions of temperatures of SAD body and front. SADs are classified into two types (SAD$_{Tgra}$, SAD$_{Timp}$) depending on whether the temperature increases significantly. \textbf{b-d}: Width distributions of two types of SADs in the whole region, in the north region, and in the south region according to the classification of body temperature evolution. \textbf{f-h}: Similar to \textbf{b-d}, but for the front temperature of SADs. }
\label{fig:fig8}
\end{figure*}

\begin{figure*}
\centering
\includegraphics [width=0.98\textwidth]{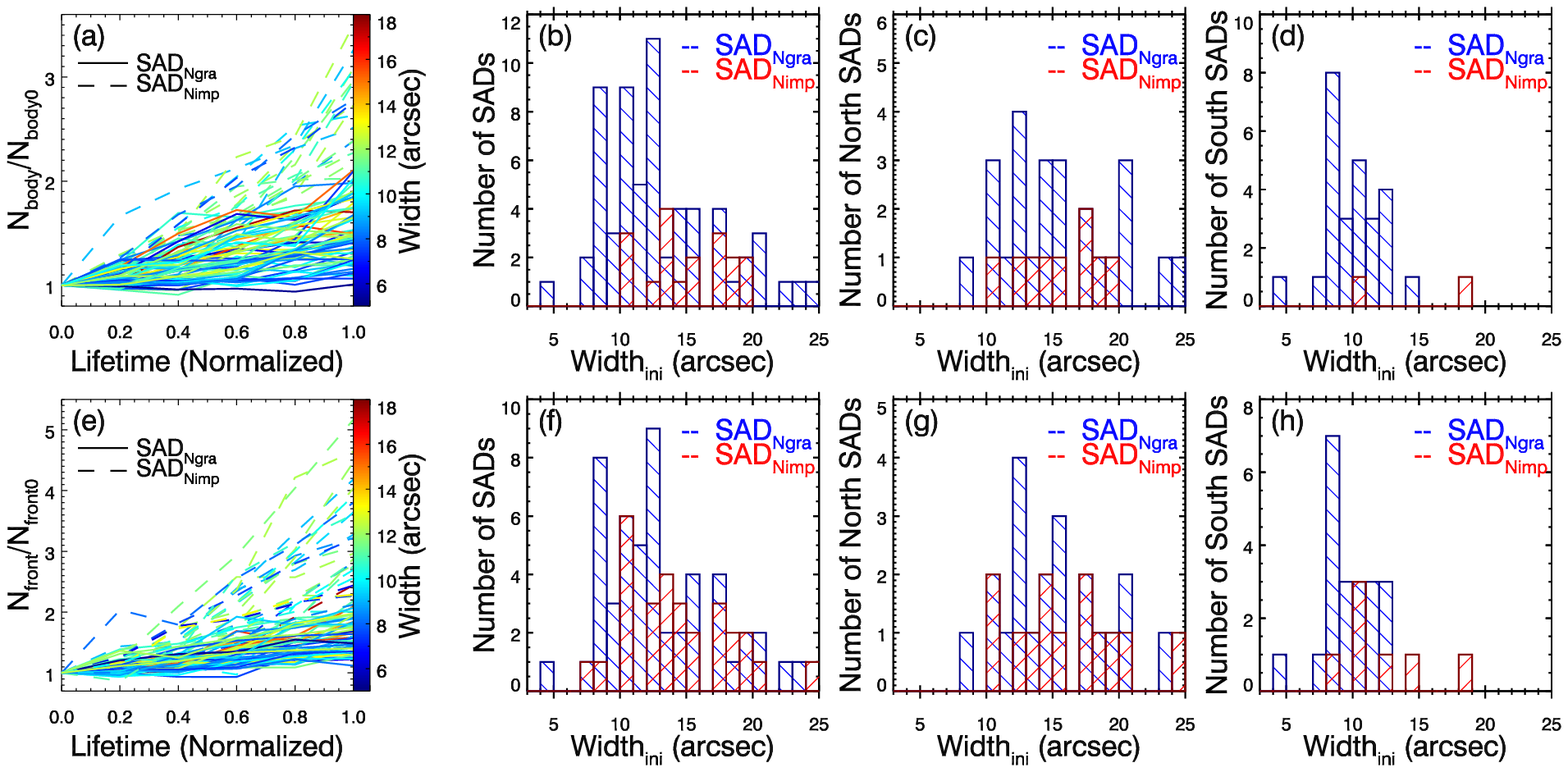}
\caption{Similar to Figure \ref{fig:fig8}, but for two types of SADs (SAD$_{Ngra}$, SAD$_{Nimp}$) classified according to the body and front density evolution properties.}
\label{fig:fig9}
\end{figure*}

\subsection{Width distributions of SADs with different thermal evolution properties} \label{subsec:DEMC}
Figures \ref{fig:fig8}(a,e) exhibit the temperature evolution of 81 SADs, and they are divided into 2 groups: SAD$_{Tgra}$ and SAD$_{Timp}$ as mentioned at the beginning of this Section. Panels (b,f) show the width distribution of two types of SADs in the whole region. It reveals that width distributions of these two types of SADs with different temperature evolution characteristics are basically similar, which is consistent with result as shown in Figures \ref{fig:fig5}(b,c). To exclude influence of the magnetic environment factor, SADs in the north and south regions are respectively shown in panels (b,d) and (g,h). It is obvious that most SADs in the south region have small initial widths and are classified as SAD$_{Tgra}$ for the body temperature, which result in that the width of SAD$_{Tgra}$ is slightly smaller than that of SAD$_{Timp}$ as shown in panel (b).

\begin{figure*}
\centering
\includegraphics [width=0.99\textwidth]{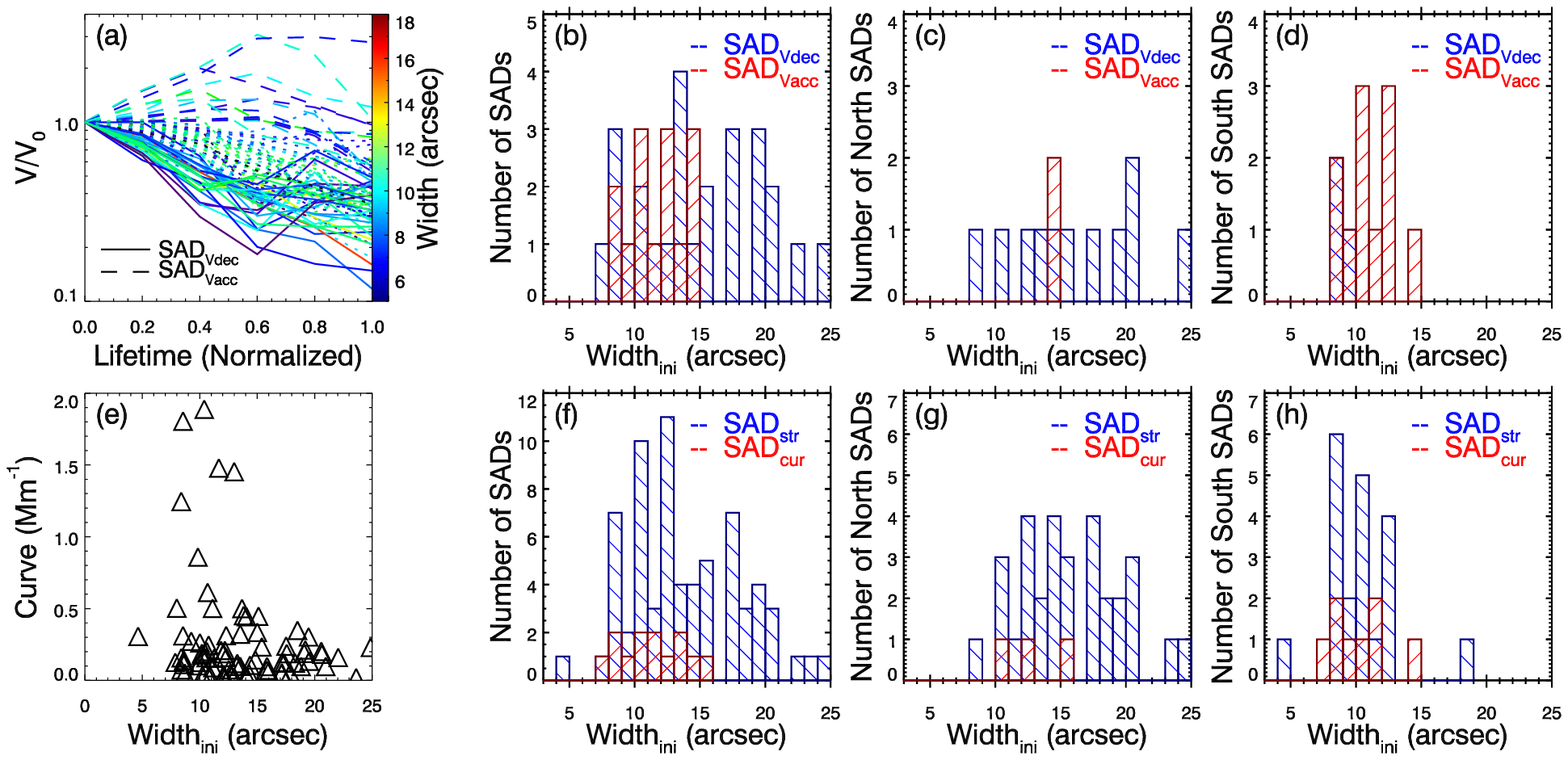}
\caption{Width distributions of SADs with different velocity evolution property and trajectory curvature. \textbf{a}: Evolutions of SAD velocity. According to the evolution of velocity, SADs are classified into two types (SAD$_{Vdec}$, SAD$_{Tacc}$). \textbf{b-d}: Width distributions of two types of SADs in the whole region, in the north region, and in the south region. \textbf{e}: Scatterplot of SAD initial width vs trajectory curvature. According to the value of trajectory curvature, SADs are classified into two types (SAD$_{str}$, SAD$_{cur}$). \textbf{f-h}: Similar to \textbf{b-d}, but for the trajectory curvature. }
\label{fig:fig10}
\end{figure*}

\begin{figure*}
\centering
\includegraphics [width=0.9\textwidth]{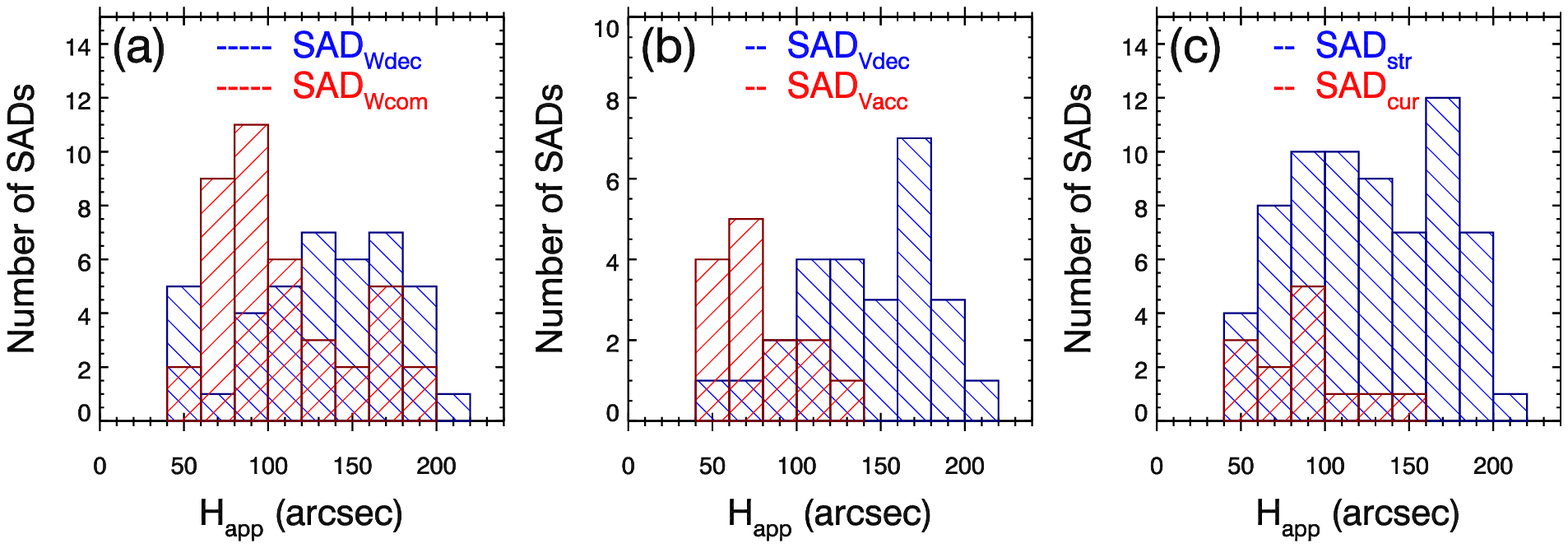}
\caption{Statistical distributions of the appearance height of two groups of SADs with different width evolution properties (\textbf{a}, SAD$_{Wdec}$ and SAD$_{Wcom}$), velocity evolution properties (\textbf{b}, SAD$_{Vdec}$ and SAD$_{Vacc}$), and trajectory curvature (\textbf{c}, SAD$_{str}$ and SAD$_{cur}$), respectively.}
\label{fig:fig10_check}
\end{figure*}

Similar analysis is also conducted for the density of SAD body and front. As shown in Figure \ref{fig:fig9}, two groups of SADs with different density evolution characteristics shows no remarkable difference in their width distributions, which is consistent with Figures \ref{fig:fig5}(e,f). Here one can see that no significant differences of initial width distribution are found between the two types of SADs with different evolution characteristics of thermal parameters like temperature and density, indicating the same formation mechanism of these SADs.

\subsection{Width distributions of SADs with different kinematic evolution properties} \label{subsec:DEMK}
We further investigated the initial width distributions of SADs with different velocity evolution properties and trajectory curvatures. Figure \ref{fig:fig10}(a) exhibits the velocity evolution of 81 SADs, which could be divided into 2 groups: SAD$_{Vdec}$ and SAD$_{Vacc}$ as mentioned at the beginning of this Section. As shown in panels (b,c), SAD$_{Vacc}$ tend to have smaller initial widths than SAD$_{Vdec}$, which is consistent with Figure \ref{fig:fig5}(d). By checking appearance heights of these SADs, we found that the appearance heights of SAD$_{Vacc}$ are overall smaller than the appearance heights of SAD$_{Vdec}$ as shown in Figure \ref{fig:fig10_check}(b), where more frequent collisions between SADs would happen and result in intermittent acceleration of SADs. Figures \ref{fig:fig10}(c,d) also reveal that SADs in the north region are mostly decelerated, while the ones in the south region seem to be mainly accelerated. It may be due to the fact that the appearance height of the south SADs is obviously lower than that of the north SADs, and the SADs with lower appearance heights are more prone to acceleration (as shown in Figure \ref{fig:fig10_check}(b)).

Figures \ref{fig:fig10}(e)-(g) reveal that there are some SADs with large trajectory curves, and they all have small initial widths. It is also found that the appearance heights of SAD$_{cur}$ are overall smaller than the appearance heights of SAD$_{str}$ (Figure \ref{fig:fig10_check}(c)). Therefore, we speculate that at lower height, the interactions like collisions or merging between different SADs, and interactions with surrounding turbulent plasma, could take place more frequently and affect the trajectories of SADs.

Although beyond the scope of this study, we want to note that in the future, more cases of merging or splitting SADs should be investigated in detail with high-resolution observations. Comprehensive analyses of heights, widths, velocities, acceleration rates, and trajectory curvatures of these typical SADs at different times during their descents and the correlations between these properties are very important to understand SADs' dynamic behaviors and the underlying evolution mechanism.

\section{Summary} \label{sec:Con}
In this study, we statistically investigated widths of 81 SADs observed during the flare occurring on May 22, 2013. After measuring widths of these SADs during their descents, we obtained their width distributions and evolution characteristics of widths. Then for the first time, we conducted correlation analysis between width and the other parameters of these SADs. Our findings can be summarized as follows:
\begin{enumerate}
\item All SAD widths measured at 6 moments during the descent show a log-normal distribution, while the initial width, maximum width and average width of SADs during the descent show a double-peak structure in this flare. We find that this double-peak structure is caused by the fact that SADs observed in the north flare region tend to have larger observed widths than those in the south region.
\item Combining SDO and STEREO-A images, we reconstructed 3D structures of flare loops below the SADs and found that the flare loops in the north have a larger angle to the line of sight than those in the south, which may affect the observed widths of SADs in the two regions.
\item By comparing the properties of plasma surrounding the SADs as well as their heights in the north and south regions, we find that the SADs in the south region appear in a relatively lower altitude thus have a plasma environment with higher density, which tends to breed SADs with small widths.
\item Excluding the effect of magnetic field and plasma environment, SADs with different initial widths show no significant differences in their appearance time, disappearance heights, and initial projective velocities. However, SADs with larger initial widths usually have larger appearance heights, lifetimes and thus lower temperatures and densities than those with smaller initial widths.
\item Most SADs with large initial widths show roughly monotonous width decelerations during their descents while widths of some SADs with small initial widths could show complicated evolutions. We suggest that former type of SADs tend to appear at a higher height and are nature to show obvious decrease in their spatial scale during the descent to lower atmosphere with stronger magnetic fields.
\item No significant differences of initial width distribution are found between the two types of SADs with different temperature or density evolution characteristics, indicating that these SADs are produced by same mechanism.
\item SADs with obvious acceleration or curved trajectories tend to have smaller initial widths and heights. We suggest that it is caused by the fact that SADs with small initial widths usually appear in lower heights, where more frequent collisions could lead to the intermittent acceleration and curved trajectories.
\end{enumerate}

Based on the results reported in the present study, we can now give answers to those questions mentioned at the end of Introduction. In general, all SAD widths measured at 6 moments during the descent show a log-normal distribution, indicating random and unstable processes involved in the formation and evolution of SADs. And magnetic environment is one key factor determining widths of SADs as well as their evolution characteristics. For example, in the flare reported here, the difference of magnetic environments in the north and south regions results in the width difference of SADs observed in the two regions.
The SADs with different initial widths show no significant differences in their temperature and density evolution characteristics. But for the SADs with small initial widths, they usually appear in lower heights, where more frequent collisions could lead to their intermittent acceleration, width increment, and curved trajectories. According to these results, we propose that SADs with different initial widths seem to be produced by same formation mechanism while various external factors such as the collision and mergence between SADs, plasma conditions around SADs (like density), and the magnetic field configuration could affect their subsequent width evolutions.

\section*{Acknowledgements}
The authors are cordially grateful to the anonymous referee for his/her constructive comments and suggestions improving this paper.
The data used here are courtesy of the \emph{SDO} and \emph{STEREO} science teams. \emph{SDO} is a mission of NASA's Living With a Star Program. The authors were supported by the National Key R\&D Program of China (2019YFA0405000, 2022YFF0503800), the National Natural Science Foundation of China (11825301, 12273060, 12222306, 11873059, and 11903050), the Strategic Priority Research Program of the Chinese Academy of Sciences (XDB41000000), the Youth Innovation Promotion Association CAS (2023063), the Open Research Program of Yunnan Key Laboratory of Solar Physics and Space Science (YNSPCC202211), and Yunnan Academician Workstation of Wang Jingxiu (No. 202005AF150025).

\section*{Data availability}
The data underlying this article will be shared on reasonable request to the corresponding author.



\bibliographystyle{mnras}
\bibliography{SADs_paper} 

\onecolumn
\begin{ThreePartTable}
\begin{TableNotes}
\footnotesize
\item [a] According to the evolution characteristic of width, ``dec'' represents that the SAD performs roughly monotonous decrease of width
during the descent while ``com'' represents that the SAD performs complicated evolution of width.
\item [b] According to the evolution characteristic of body and front thermal properties (temperature and density), ``gra'' represents that the corresponding thermal property of the SAD increases gradually while ``imp'' represents that the corresponding thermal property of the SAD increases impulsively.
\item [c] According to the evolution characteristic of velocity, ``dec'' represents that the SAD performs great decrease of velocity at middle time while ``acc'' represents that the SAD has a significant acceleration.
\item [d] According to the trajectory curvature, ``cur'' represents that the trajectory of the SAD is curved while ``str'' represents that the trajectory of the SAD is nearly straight.
\item [e] ``/'' represents the SAD does not meet any of the two grouping criteria.
\end{TableNotes}

\begin{longtable}{ccccccccccc}
\insertTableNotes  
\endlastfoot
\caption{Different Physical Properties of the 81 SADs. \label{tab:SADs}}\\
\hline\hline
SADs & $w_{ini}$ & $w_{max}$ & Region & Width   & T$_{body}$  & T$_{front}$  & N$_{body}$  & N$_{front}$  & Velocity     & Trajectory \\
     & (arcsec)   & (arcsec) & South\textbf{\&}North    & dec\textbf{\&}com\tnote{a} & gra\textbf{\&}imp\tnote{b} & gra\textbf{\&}imp\tnote{b} & gra\textbf{\&}imp\tnote{b} & gra\textbf{\&}imp\tnote{b} & dec\textbf{\&}acc\tnote{c} & cur\textbf{\&}str\tnote{d} \\
\hline
1  & 14.4 & 17.4 & North & com & imp & imp & gra & gra & acc   & str \\
2  & 15.4 & 15.8 & North & dec & imp & imp & gra & gra & /     & str \\
3  & 10.9 & 10.9 & /\tnote{e}     & com & gra & gra & gra & gra & dec   & str \\
4  & 17.3 & 18.1 & North & dec & imp & imp & imp & imp & /     & str \\
5  & 14.3 & 14.3 & North & dec & gra & imp & gra & imp & acc   & str \\
6  & 10.8 & 10.8 & North & dec & imp & imp & imp & imp & dec   & str \\
7  & 17.2 & 17.2 & North & dec & imp & imp & imp & imp & dec   & str \\
8  & 14.8 & 14.8 & North & dec & imp & imp & imp & imp & dec   & str \\
9  & 12.0 & 12.2 & North & com & imp & imp & imp & imp & dec   & str \\
10 & 13.3 & 13.3 & /     & dec & imp & imp & imp & imp & dec   & str \\
11 & 17.6 & 17.6 & North & dec & imp & gra & gra & gra & /     & str \\
12 & 18.3 & 18.3 & North & dec & imp & imp & imp & imp & /     & str \\
13 & 16.0 & 18.1 & /     & dec & gra & imp & imp & imp & dec   & str \\
14 & 11.8 & 13.4 & North & com & gra & imp & gra & gra & /     & str \\
15 & 15.8 & 15.8 & North & com & gra & gra & gra & gra & /     & str \\
16 & 15.9 & 15.9 & /     & com & gra & gra & gra & gra & /     & str \\
17 & 19.1 & 19.1 & North & com & imp & imp & imp & imp & dec   & str \\
18 & 17.5 & 17.5 & /     & com & imp & imp & imp & imp & dec   & str \\
19 & 12.6 & 17.3 & North & com & gra & gra & gra & gra & /     & str \\
20 & 13.2 & 14.4 & North & dec & gra & imp & imp & imp & /     & str \\
21 & 8.6  & 8.6  & North & com & gra & gra & gra & gra & dec   & str \\
22 & 12.1 & 12.1 & /     & dec & gra & gra & gra & gra & /     & str \\
23 & 13.0 & 13.0 & North & dec & gra & imp & gra & gra & /     & cur \\
24 & 10.6 & 22.8 & South & com & imp & imp & imp & imp & /     & str \\
25 & 24.9 & 24.9 & North & dec & gra & imp & gra & imp & dec   & str \\
26 & 8.0  & 8.0  & South & dec & gra & imp & gra & gra & /     & cur \\
27 & 12.3 & 12.9 & South & com & gra & gra & gra & gra & acc   & str \\
28 & 10.2 & 10.2 & North & com & gra & gra & gra & gra & /     & str \\
29 & 13.5 & 13.5 & North & dec & gra & gra & gra & gra & dec   & str \\
30 & 23.6 & 23.6 & North & dec & gra & gra & gra & gra & /     & str \\
31 & 17.1 & 17.1 & North & com & gra & imp & gra & gra & /     & str \\
32 & 21.0 & 21.0 & North & dec & gra & gra & gra & gra & /     & str \\
33 & 13.0 & 13.0 & South & dec & gra & gra & gra & gra & acc   & str \\
34 & 15.0 & 19.8 & North & com & gra & imp & imp & imp & /     & str \\
35 & 18.8 & 18.8 & South & dec & gra & imp & imp & imp & /     & str \\
36 & 14.1 & 14.1 & North & dec & gra & gra & gra & gra & /     & str \\
37 & 9.8  & 12.4 & South & com & gra & gra & gra & gra & /     & str \\
38 & 12.5 & 12.5 & North & dec & gra & gra & gra & gra & /     & str \\
39 & 10.2 & 11.5 & North & com & gra & gra & gra & gra & /     & str \\
40 & 8.4  & 9.1  & South & com & gra & gra & gra & gra & acc   & str \\
41 & 20.6 & 20.6 & North & dec & gra & imp & gra & imp & dec   & str \\
42 & 8.4  & 8.7  & South & com & gra & gra & gra & gra & dec   & cur \\
43 & 18.5 & 18.5 & North & dec & gra & gra & gra & gra & /     & str \\
44 & 12.3 & 15.3 & North & com & gra & gra & gra & gra & /     & str \\
45 & 12.3 & 12.3 & South & com & imp & imp & gra & imp & /     & str \\
46 & 19.5 & 19.5 & North & dec & gra & gra & gra & gra & /     & str \\
47 & 20.6 & 20.6 & North & dec & gra & gra & gra & gra & dec   & str \\
48 & 10.4 & 10.4 & North & dec & imp & imp & gra & imp & /     & cur \\
49 & 11.2 & 11.2 & /     & com & gra & gra & gra & gra & dec   & str \\
50 & 13.7 & 15.6 & /     & dec & gra & imp & imp & imp & acc   & cur \\
51 & 10.0 & 10.0 & /     & com & imp & imp & imp & imp & /     & str \\
52 & 9.8  & 11.5 & South & com & gra & imp & gra & gra & dec   & cur \\
53 & 19.2 & 19.2 & /     & com & imp & imp & imp & imp & dec   & str \\
54 & 15.1 & 15.1 & North & dec & gra & gra & gra & gra & dec   & cur \\
55 & 8.5  & 10.6 & South & com & gra & gra & gra & gra & /     & str \\
56 & 13.8 & 13.8 & /     & dec & imp & imp & imp & imp & dec   & cur \\
57 & 7.9  & 8.0  & /     & com & imp & imp & gra & imp & dec   & str \\
58 & 11.1 & 11.1 & South & com & gra & gra & gra & gra & acc   & str \\
59 & 8.4  & 8.4  & South & dec & gra & imp & gra & gra & /     & str \\
60 & 10.6 & 16.1 & South & com & gra & imp & gra & imp & acc   & str \\
61 & 10.7 & 10.7 & South & dec & gra & gra & gra & gra & acc   & cur \\
62 & 10.0 & 10.0 & South & com & gra & gra & gra & gra & acc   & str \\
63 & 10.2 & 14.8 & South & com & gra & imp & gra & imp & /     & str \\
64 & 12.1 & 12.1 & South & dec & gra & gra & gra & gra & acc   & str \\
65 & 8.6  & 9.9  & South & dec & gra & gra & gra & gra & /     & str \\
66 & 10.6 & 10.6 & South & com & gra & imp & gra & gra & /     & str \\
67 & 9.3  & 10.1 & South & com & gra & imp & gra & gra & acc   & str \\
68 & 11.1 & 16.0 & South & com & gra & gra & gra & gra & /     & cur \\
69 & 22.0 & 22.0 & /     & dec & gra & gra & gra & gra & dec   & str \\
70 & 8.7  & 10.0 & South & com & gra & gra & gra & gra & acc   & str \\
71 & 8.5  & 9.3  & South & com & gra & gra & gra & gra & /     & cur \\
72 & 13.5 & 13.5 & /     & dec & gra & gra & gra & gra & dec   & str \\
73 & 14.1 & 15.9 & South & com & gra & imp & gra & imp & acc   & cur \\
74 & 12.2 & 13.2 & /     & dec & gra & gra & gra & gra & /     & str \\
75 & 4.6  & 6.4  & South & com & gra & gra & gra & gra & /     & str \\
76 & 11.7 & 11.7 & South & com & gra & imp & gra & gra & /     & cur \\
77 & 8.6  & 10.2 & South & com & gra & imp & gra & imp & dec   & str \\
78 & 17.9 & 17.9 & /     & dec & gra & gra & gra & gra & dec   & str \\
79 & 19.5 & 19.5 & /     & dec & gra & gra & gra & gra & dec   & str \\
80 & 12.2 & 12.2 & /     & dec & gra & gra & gra & imp & /     & str \\
81 & 17.1 & 17.1 & /     & dec & gra & gra & gra & gra & /     & str \\
\hline
\end{longtable}

\end{ThreePartTable}

\twocolumn

\end{document}